# Designing 1D correlated-electron states by non-Euclidean topography of 2D monolayers

Sunny Gupta[1], Henry Yu[2], and Boris I. Yakobson[1,3,4*]

[1]Department of Materials Science and Nanoengineering, Rice University, Houston, TX, 77005 USA
[2]Applied Physics Program, Rice University, Houston, TX, 77005 USA
[3]Department of Chemistry, Rice University, Houston, TX 77005, USA
[4]Smalley-Curl Institute for Nanoscale Science and Technology, Rice University, Houston, TX, 77005 USA

*Email: biy@rice.edu

**Abstract**

Two-dimensional (2D) bilayers, twisted to particular angles to display electronic flat bands, are being extensively explored for physics of strongly correlated 2D systems. However, the similar rich physics of one-dimensional (1D) strongly correlated systems remains elusive as it is largely inaccessible by twists. Here, a distinctive way to create 1D flat bands is proposed, by either stamping or growing a 2D monolayer on a non-Euclidean topography-patterned surface. Using boron nitride (hBN) as an example, our analysis employing elastic plate theory, density-functional and coarse-grained tight-binding method reveals that hBN's bi-periodic sinusoidal deformation creates pseudo- electric and magnetic fields with unexpected spatial dependence. A combination of these fields leads to anisotropic confinement and 1D flat bands. Moreover, changing the periodic undulations can tune the bandwidth, to drive the system to different strongly correlated regimes such as density waves, Luttinger liquid, and Mott insulator. The 1D nature of these states differs from those obtained in twisted materials and can be exploited to study the exciting physics of 1D quantum systems.

Recently, twisted bilayer graphene (TBG) at magic angles[1,2] and other van der Waals (vdW) heterostructures at small twist[3,4] have garnered great attention as material platforms for realizing 2D correlated physics with an unprecedented level of control. Several interesting electronic phases have been observed in these systems, such as correlated insulator[2,4], superconductivity[1,5], non-trivial electronic topology[6], and magnetism[7,8]. Physically, these emergent phases can be attributed to the existence of the Bloch flat bands[9,10], where the kinetic energy scale is quenched, and the role of electronic interactions is enhanced. The flat bands originate from the perturbation of the electronic structure by the long-wavelength superlattice (moiré) period, which suppresses the group velocity in TBG at magic angles[11,12] and creates electronic confinement in other vdW heterostructures[13–15]. The moiré periods arise from either lattice mismatch or rotational misalignment between the layers with fine-tuning of the twist, posing challenges[16] in fabrication, variability between devices and scalability. Moreover, examining the rich unexplored physics of 1D strongly correlated systems, likewise, is largely inaccessible by twists.

Below we describe creating flat bands through an alternative route, not requiring a twist angle. The strategy involves either growing[17] or stamping[18] a 2D material on a topographically patterned substrate with non-zero Gaussian curvature, that is non-Euclidean surface, Fig. 1a. To conform to such surfaces, a planar 2D crystal must deform, so the undulated topography imparts strain. Strain is known to perturb the crystal Hamiltonian through a deformation potential[19], to a magnitude proportional to the strain, which in turn is determined by the topography of the surface. A periodic strain modulation will create a confining potential, which --if strong enough--can localize electrons and result in modulated super-lattice band whose bandwidth depends on the surface's specific geometry. Hence, patterned surfaces with specific topography can, in general, create and fine-tune flat bands in any 2D semiconductor materials. To the best of our knowledge, creating either 2D or 1D flat bands in monolayer semiconductors by undulation has not been discussed before.

We illustrate this idea by theoretically investigating the electronic properties of monolayer hexagonal boron nitride (hBN) deformed by a bi-periodic sinusoidally modulated topography. Interestingly, strained hBN attains both pseudo-electric field $E_P$ (by virtue of deformation potential) and also (having a honeycomb lattice, like graphene, with two inequivalent basis atoms) pseudo magnetic field $B_P$[20]. We find that for bi-sinusoidal deformation, $E_P$ and $B_P$ have very different spatial dependence. A combination of these leads to anisotropic confinement and creates one-dimensional (1D) flat bands, whose bandwidth can be varied by the surface topography. The 1D nature of these states can be exploited to probe the exciting physics of one-dimensional quantum systems, which has been predicted to exhibit interesting effects[21] such as Luttinger liquid behavior, charge and spin density waves, Peierls instability, and deviation from Fermi-liquid theory. The origin and nature of these 1D states are different from the 2D flat bands observed in TBG and twisted vdW heterostructures, opening an exciting realm of exploring many-body effects in 1D quantum systems in a clean and controllable manner.

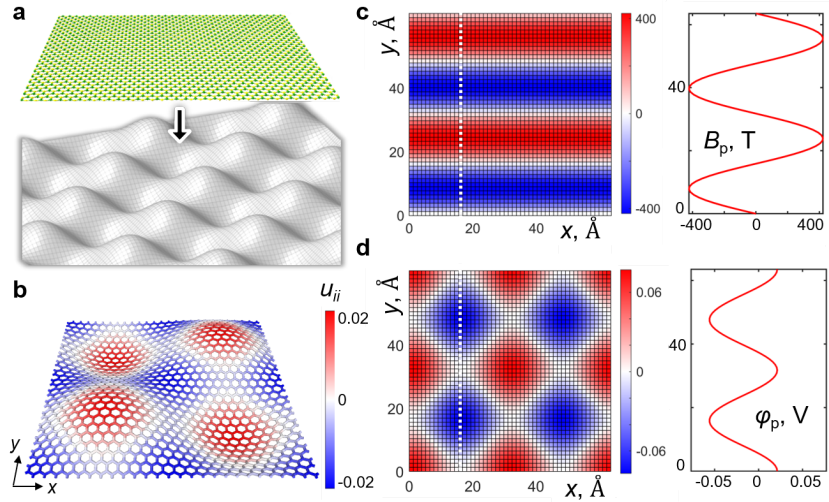

**Fig. 1. Design scheme to create flat bands using mechanical deformation induced pseudo-electric and magnetic fields.** (a) A 2D material stamped on a topography with bi-periodic sinusoidal height-modulation, causing in-plain strain. (b) Relaxed h-BN geometry and strain field $\varepsilon \equiv u_{ii}$ at undulation aspect ratio A = 0.079. The left panels in (c) and (d) map the pseudo-magnetic ($B_P$) and pseudo-electric potential ($\varphi_P$) fields, respectively; the right panels show the values of the respective fields along $x$ = const, marked by a white dashed line on the left.

A 2D material conformed to a curved non-Euclidean surface undergoes a locally in-plain strain, which can be evaluated at a continuum level (for relatively smooth topography) by solving the second Föppl–von Kármán (FvK) equation[22], $\Delta^2 \chi = -Y(f_{xx}f_{yy} - f_{xy}^2)$.

Here $\chi$, $Y$, and $f$, are the Airy stress function, Young's modulus, and the surface shape function, respectively. We consider bi-periodic sinusoidally modulated off-plane shape $f(x,y)$ = $h\sin\alpha x \cdot \sin\alpha y$, with $\alpha \equiv 2\pi/L$, akin to *egg-cart*. The displacement ($u_x$, $u_y$) and imparted strain ($u_{ij}$) fields were solved analytically from the FvK equation for the sinusoidal surface (for details see Supplementary Information, SI-1). The maximum tension or compression for this geometry depends on the aspect ratio A = $h/L$, as $\varepsilon = |u_{ii,max}| = \pi^2 A^2 (1-\nu)/2$, where $\nu = 0.31$ is the Poisson's ratio for hBN and $u_{ii}=(u_{xx}+u_{yy})/2$. A relaxed atomic structure of a sinusoidal hBN can be further constructed from the solved displacements $u_x$ and $u_y$, as in Fig. 1b, for A=0.079, where color shows $u_{ii}$ (for details see SI-3). As expected, the hilltops and valley-bottoms are stretched while the saddle areas are compressed, to an amplitude $\varepsilon$~2.12%. We note here that, generally, the material strain depends on both the surface shape $f(x,y)$ and boundary conditions (in case of growth, also on the chemical potential $\mu_{hBN}$, controlled by the growth conditions). Here we allow full relaxation to minimise elastic energy while accommodating the substrate topography, that is no forces at (remote) layer's perimeter and no friction to the substrate. This corresponds to "stamping" the 2D material onto the frictionless matrix, when the layer contracts laterally, with non-negligible displacements $u_x$, $u_y$ (see SI-1).

Analogous to graphene[23–26], the strain in hBN generates pseudo- electric and magnetic fields[20,23], significantly perturbing the crystal Hamiltonian. The low energy effective Hamiltonian in strained hBN in the vicinity of the K points is given by

$$H^{(\tau)} = \hbar v_F \sigma^{(\tau)} \cdot (k - \tau A_p) - e\varphi_p \quad (1)$$

where $v_F = 3|t|a/2$, $|t|$ is the nearest-neighbor (NN) hopping amplitude, and $a$ is the interatomic distance. $\sigma^{(\tau)}=(\tau\sigma_x,\sigma_y,\sigma_z)$ are defined in terms of the three Pauli matrices, and $\tau=+1$ (-1) for K (K'). $k = (k_x, k_y, \Delta)$, where $\hbar v_F \Delta$ is the difference in sublattice potential between B and N atoms, and $k_{x,y}$ is the electron crystal momentum measured relative to K or K'. $A_p$ is the pseudo-vector potential caused by shear, $A_p = (\beta_0/\sqrt{2}a)[(u_{xx}-u_{yy})/2, -u_{xy}]$, where $\beta_0 = (a/t)\partial t/\partial a$ = -3.3. $\varphi_p$ is the pseudo-electric potential (PEP) arising due to the hydrostatic component of strain, $\varphi_p = -gu_{ii}$, where $g \approx 3.66$ V[27]. Accordingly, these potentials generate pseudo-electric field (PEF) $E_p = -\nabla\varphi_p$ and pseudo-magnetic field (PMF) $B_p = B_p z$, where $B_p = (\hbar/e)(\partial_x A_{p,y} - \partial_y A_{p,x})$ and $z$ the unit z-vector. One can already recognize that these additional pseudo fields in the Hamiltonian, arising due to strain, act as a perturbing confinement potential.

The strain fields obtained for sinusoidal surfaces allow us to derive the analytical expressions for pseudo electric and magnetic fields (for details see SI-2).

$$\varphi_p = g((1-\nu)/16)h^2\alpha^2(\cos2\alpha x + \cos2\alpha y) \quad (2a)$$
$$B_p = -(\beta_0\hbar/\sqrt{2}ae)((1+\nu)/8)h^2\alpha^3\sin2\alpha y \quad (2b)$$

Fig. 1c,d shows the PMF and PEP for A=0.079, and L=6.35 nm. The spatial dependence for both fields is different and surprisingly, PMF depends only on $y$ (Fig. 1c). It is known[28] that periodic magnetic fields can lead to confinement and create localized electronic states. Similarly, we expect that for sinusoidally modulated hBN, a combination of both PEP and PMF will create flat bands. Sections of these fields along y-direction, at x = const are plotted in the right panels of Fig. 1c,d. The periodic PMF has an amplitude of $B_{p,max}$~420 T, which corresponds to confinement energy of ~$2\mu_B B_{p,max}$ = 49 meV, while the periodic PEP corresponds to confinement energy of ~76 meV. We will show that different spatial dependence of PMF and PEP leads to anisotropic confinement and results in the interesting electronic nature of the flat bands.

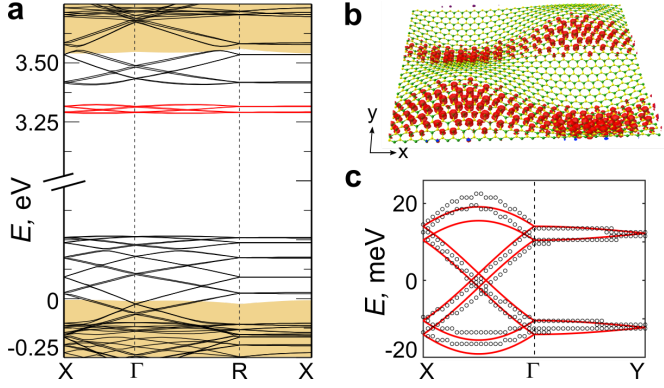

**Fig. 2. Electronic structure of modulated hBN.** (a) Calculated electronic band structure of sinusoidally deformed hBN, A = 0.079. The marked red defect-like flat bands are due to electronic confinement caused by deformation. Yellow-shaded areas mark the bands of pristine hBN with gap in between. (b) The band decomposed charge density corresponding to one of the flat bands at Γ point $|\Psi_{nk}|^2$ shows the one-dimensional electronic nature of the bands. (c) An enlarged view of the flat bands, red in (a), plotted along X-Γ-Y. The solid red line is the fit with a 8-band tight-binding model.

We next calculate the electronic bands of our topographically-strained hBN, using density-functional based tight-binding (DFTB) theory with a local orbital basis[29]. DFTB has been successfully applied to study various forms of hBN[15,30], for which DFT calculations are intractable (see SI-4 for details).

Monolayer hBN honeycomb lattice is akin to graphene, yet the different basis atoms break the sub-lattice A-B symmetry, and an energy gap opens, making hBN an insulator. The undeformed monolayer hBN shows a band gap of ~3.55 eV. Fig. 2a shows the band structure under bi-sinusoidal strain, with A = 0.079, $\varepsilon$ = 2.12%, and $L$ = 6.35 nm. One can see additional bands appearing in the gap, looking like defect states which might arise due to electronic confinement. We find that the bandwidth ($W$) of the lowest unoccupied states (shown in red) is $W$ = 39 meV, which is very small, and it is a flat band; in comparison, the effective $W$ of pristine BN bands corresponding to NN hopping $t$~2.16 eV[20] is $W$~$4t$=8.6 eV, which is much larger than the $W$ of the modulated flat bands. These flat bands are well separated by >100 meV from the other states at higher energies. Interestingly, the bands are dispersive along Γ-X and almost non-dispersive along R-X, which corresponds to $k_x$ and $k_y$ directions, respectively. This makes these flat bands one-dimensional and very different from those seen in TBG and other twisted vdW heterostructures. The band decomposed charge density $|\Psi_{nk}|^2$ in Fig. 2b corresponds to one of the flat bands at Γ point. The electronic states are delocalized along the $x$- but are completely localized along the $y$-direction, confirming these flat bands' 1D nature.

The flat bands are composed of 8 electronic states and are localized mainly on the four extremes of the sinusoidal modulation in Fig. 2b. Fig. 2c shows the enlarged view of the flat bands plotted along X-Γ-Y. We find that the flat bands dispersion and charge modulation can be described by a simple "coarse-grained" 8-band tight-binding (TB) Hamiltonian,

$$H = \Sigma_x t_x c^\dagger_{x,y} c_{x+2,y} + \Sigma_y t_y c^\dagger_{x,y} c_{x,y+1} \qquad (3)$$

where $t_x$ and $t_y$ are hopping amplitudes along $x$-, and $y$-direction, respectively (for details see SI-5). The 8 bands arise from the states localized on the 2 maxima and 2 minima, and each of them being doubly occupied. The maxima and minima act as artificial "quantum dots". Fitting this TB model to DFTB results gives $|t_x|$ = 8.9 meV (~$W$/4) and $|t_y|$ = 0.9 meV. The ratio of hopping along $x$-, and $y$-direction is $|t_y|/|t_x|$ = 0.1, again a manifestation of one-dimensionality of the electronic states. This is quite surprising at first, because the strain pattern appears to be isotropic along the *x* and *y*-direction (Fig. 1b).

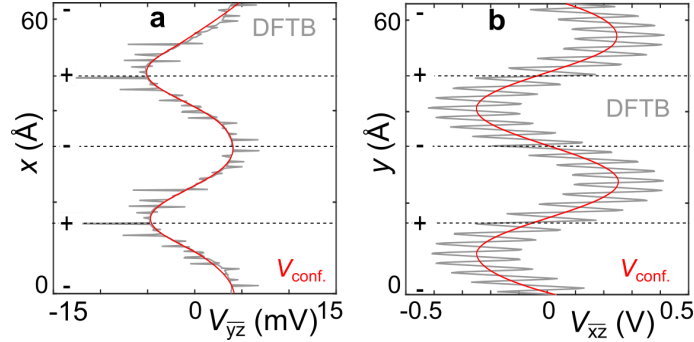

**Fig. 3. Total and long-range modulated average electrostatic potential.** Electrostatic potential along (a) *x*, and (b) *y* and averaged along the other two perpendicular directions. The solid grey lines correspond to the calculated averaged potential, while the thin red lines correspond to the long-range potential modulation extracted by Gaussian averaging. The + and - mark tensile and compressive regions.

To gain microscopic insights into the reasons behind these flat bands' 1D nature, we calculated the electrostatic potential along *x* (Fig. 3a) or *y* (Fig. 3b) while averaged along the other two perpendicular directions. The sharp features are due to the approximations used to evaluate the diverging potential near atomic sites (for details see SI-4). The potential rapid oscillations are due to periodic atomic sites, but a long-range modulation can also be seen. Gaussian averaging extracts the long-range modulation ($V_{conf.}$, red lines in Fig. 3a-b), which is very different along *x*- and *y*-directions: e$V_{conf.}$ along *y*-direction has a depth of ~500 meV, larger than mere ~9 meV along *x*. This anisotropic confinement is expected because of the different spatial dependence of PEP and PMF, Fig. 1c-d. The smaller $V_{conf.}$ along *x*-direction is mainly due to contributions from PEP only, while the larger $V_{conf.}$ along the *y*-direction is contributed by both PEP and PMF. This signifies that the long-range potential modulations are due to pseudo-electric and magnetic fields, providing the anisotropic confinement needed to maintain one-dimensional flat bands. Additionally, the larger confinement energy along *y*-direction results in lower hopping amplitude $t_y$ in our coarse-grained model.

We find that the width of these flat bands can be tuned by changing either the aspect ratio of the topography, to alter the strain, or the period *L* of the undulation. Fig. 4 shows the variation of *W* as a function of *L*, and *ε*. *W* goes below 10 meV for *L*>8 nm. Moreover, *W* is found to depend exponentially on both *L*, and *ε*, a good fit $W \propto \exp(-0.56*L - 0.34*\varepsilon)$ is obtained (blue surface in Fig. 4); this is expected since *W* effectively corresponds to the hopping amplitude between the coarse-grained sites, defined by hybridization/overlap of wave function between them. Since the strength of hybridization decreases exponentially with distance *L*, *W* is found to show the same dependence. Additionally, the exponential dependence of *W* on *ε* can also be understood from the well-known dependence of hopping energy (*t*) with strain[20], $t(a)=t_0\exp(-|\beta_0|(a/a_0-1))$, is the hopping amplitude at the effective bond length *a*, and ($a/a_0$-1) is its strain. We point out that flat bands appear even at strain as low as ~1.5%, and these bands are well separated by >70 meV from the other unoccupied bands above, which makes them accessible to experiments.

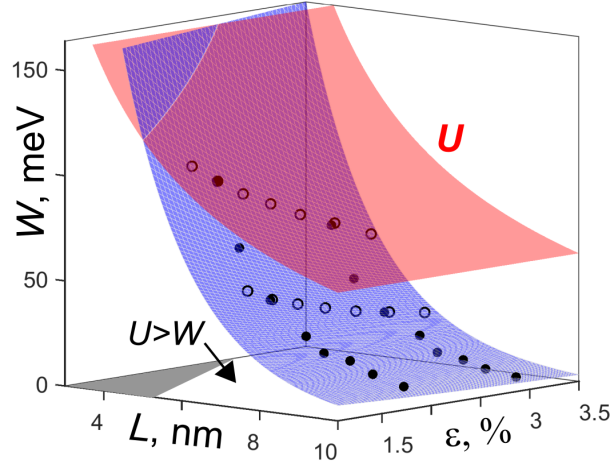

**Fig. 4. Tuning bandwidth by changing topography of undulations.** The flat band width $W$ versus topography period $L$ and strain $\varepsilon$. The blue surface shows the fitted expression $W \propto \exp(-0.56*L - 0.34*\varepsilon)$, while the empty and filled circles are computed values. The red surface shows the estimate of the on-site Coulomb energy $U$, which depends on period $L$. In the ($L$, $\varepsilon$) parameter-plane, the outside of a shaded grey is the $U > W$ region where physics of strongly correlated 1D phases can be realized.

Interesting strongly correlated physics in one-dimension[31] is expected when the ratio of on-site Coulomb interaction $U$ (responsible for electronic correlation) and the hopping amplitude $t$ is large, $U/4t = U/W >1$. Since $U$ depends inversely on length $L$, $U \propto 1/L$[13], and $t \propto \exp(-|\beta_0|L)$[20], the condition $U/W >1$ should be easily achievable for reasonable $L$. We estimate $U$ for our systems as $U=e^2/2\pi\kappa L$[13], where $L$ is the length of periodic modulation, and $\kappa = 4.73$[32] is the effective dielectric constant of hBN. The red surface in Fig. 4 shows values of $U$. $U/W$ can be enhanced by either increasing the aspect ratio (strain) and/or increasing the periodic length (Fig. 4). The region in ($L$, $\varepsilon$) with $U<W$ is shown by the grey shaded area in Fig. 4. To achieve strongly correlated phases, where $U>W$, topography with $L$ and $\varepsilon$ values lying outside of the grey area in Fig. 4 must be chosen. Depending on the ratio of $U/W$ and band filling, one can expect[31] different phases such as Mott insulator (MI), Luttinger liquid, bond ordered wave (BOW), and band insulator (BI). E.g., at small band filling, gradually changing $U/W$ from 0 to 2 can change the electronic phases in order BI → BOW → MI. Accordingly, one should expect that changing the periodic topography will provide a unique control to drive the system to different strongly correlated regimes exhibiting interesting physics. To realize the strongly correlated physics, the flat bands (Fig. 2a) must be partially filled, perhaps by electrostatic doping, as routinely done for 2D materials, including twisted bilayer TMDs[4], and graphene[2]. Long-range ordered quantum phases in 1D tend to get destroyed due to thermal fluctuations at finite temperature. Importantly, our predicted system with parallel 1D states (resulting from the quasi 1D flat bands in a 2D material, Fig. 2b) will suppress such fluctuations, due to the finite coupling between them[21], helping to achieve interesting physics in 1D at finite temperatures.

To realize our predictions in experiments, hBN may be overlaid or stamped (or possibly grown directly) on patterned substrate, for instance $SiO_2$, having a band gap of ~8.9 eV[33], larger than hBN---so that there are no unwanted hybridization between hBNs electronic states and the substrate. Fabricating bi-periodic sinusoidal modulation on silicon is challenging but has already been attempted[34]. Delamination from the substrate, if the desired strain level is high may be a concern; by comparing the surface pressure due to substrate with the maximal adhesion forces of hBN to $SiO_2$ with $\gamma_{ad}$~13 meV/Å$^2$ [see SI-6], we estimate that a strain up to $\varepsilon$~2.75% at A~0.09 [see SI-3b], should be sustainable in experiments.

In summary, we have shown that deforming a 2D semiconducting monolayer with a particular topography having non-zero Gaussian curvature, can be used as a unique and straightforward way to create flat bands and drive the system into different strongly correlated electronic regimes. Topographical modulation can be created by electron-beam lithography and does not require accurate fine-tuning of the twist angle and overcomes several challenges of twisted systems. For hBN as an example, we show that bi-periodic sinusoidal modulation generates pseudo-electric and magnetic fields, creating anisotropic electronic confinement and one-dimensional flat bands. Our proposed way to create 1D flat bands should be applicable to a variety of 2D systems like hBN with massive Dirac fermions. These flat bands are different from those observed with twisted bilayer graphene and other vdW heterostructures. These bands' one-dimensional nature will pave the route to study the exciting physics of strongly correlated 1D systems, thereby going beyond what's achievable with twisted materials.